\documentclass[prb,twocolumn,showpacs]{revtex4}
\usepackage{graphicx,epsfig,color}

\begin{document}

\title{Dynamical, dielectric, and elastic properties of GeTe}
\author{R.~Shaltaf$^{1,2}$, E. Durgun$^{3,4}$, J.-Y. Raty$^{5}$, Ph. Ghosez%
$^{3,4}$, and X. Gonze$^{1,2}$}
\affiliation{$^1$European Theoretical Spectroscopy Facility (ETSF)\\
$^2$Unit\'{e} Physico-Chimie et de Physique des Mat\'{e}riaux \\
Universit\'{e} catholique de Louvain Place, B-1348, Louvain-la-Neuve Belgium\\
$^3$ European Multifunctional Materials Institute (EMMI)\\
$^4$ Physique Theorique des Mat\'{e}riaux, Universit\'{e} de Li\`{e}ge (B5), B-4000 Li\`{e}ge, Belgium\\
$^5$ FRS-FNRS, D\'{e}partement de Physique, Universit\'{e} de Li\`{e}ge (B5),B-4000 Li\`{e}ge, Belgium}
  
\date{\today}

\begin{abstract}
The dynamical, dielectric and elastic properties of GeTe, a
ferroelectric material in its low temperature rhombohedral phase,
have been investigated using first-principles density functional
theory. We report the electronic energy bands, phonon dispersion
curves, electronic and low frequency dielectric tensors, infra-red
reflectivity, Born effective charges, elastic and piezoelectric
tensors and compare them with the existing theoretical and
experimental results, as well as with similar quantities available
for other ferroelectric materials, when appropriate.
\end{abstract}

\maketitle

\section{Introduction}

GeTe is an interesting material from both academic and industrial
perspectives. When alloyed with antimony, the electronic and optical
properties of GeTe get dramatically modified, due to change of the
microscopic structure from the crystalline to the amorphous phase.\cite%
{Ovshinsky,Libera,Yamada} This makes it a crucial base material in phase
change alloys used in optical storage rewritable CDs and DVDs.

Besides this technological interest, GeTe attracts more academic
oriented interest for its ferroelectric properties. At higher
temperatures, it possesses the highly symmetric , paraelectric,
rocksalt cubic structure (space group \textit{F3m} No.225). Below
a critical temperature $T_{c}$, it
stabilizes in a lower symmetry ferroelectric structure (space group \textit{%
R3m} No.160) with Ge and Te ions being displaced from ideal rocksalt sites. 
The ferroelectric transition is
characterised by the softening of a zone-centre transverse optic (TO) phonon
mode propagating in the [1 1 1] direction, and the freezing-in of a relative
displacement of the crystal sublattices.\cite{Steigmeier} Unlike other IV-VI
telluride based materials such as SnTe and PbTe which have very low $T_{c}$
being $\sim $ 140 K and 2 K respectively, GeTe has $T_{c}\sim $ 720 K, which
makes it the simplest ferroelectric material existing at room temperature,
with only two atoms per primitive cell.

Due to its interesting properties as ferroelectric and phase change
material, GeTe has been the subject of many experimental and theoretical
studies. The electronic, structural and optical properties have been
investigated in the different crystalline, liquid or amorphous phases.~\cite%
{Steigmeier,Andrikopoulos,Raty1,Rabe2,
Zein,Lebedev,Littlewoo,Chattopadhyay,Onodera,Ciucivara} The cubic phase
instability and pressure induced phase transition has been the issue of
discussion in some studies.~\cite%
{Zein,Lebedev,Littlewoo,Chattopadhyay,Onodera,Ciucivara,Rabe3} However the
dynamical, mechanical and piezoelectric properties of GeTe have been largely
left aside.

A Raman inelastic scattering study of GeTe was carried out in the early work
of Steigmeier\textit{ et al.},~\cite{Steigmeier} who found two principal peaks, a
first peak at a frequency of 98 cm$^{-1}$, attributed to a degenerate $E$
mode, and a second peak at 140 cm$^{-1}$, attributed to a $A_{1}$ mode. A
more recent experimental study, has reported values of 80 and 122 cm$^{-1}$
for the $E$ and $A_{1}$ modes respectively.~\cite{Andrikopoulos}

Calculations of dynamical properties of GeTe, using density functional
perturbation theory (DFPT) were reported by Zein\textit{ et al.},\cite{Zein} who
studied the GeTe in the high temperature rocksalt structure.
They found soft phonon at $\Gamma $ and reported a value of $\sim $ 10.2$e$
for the Born effective charge. Dielectric properties of the rocksalt
structure have also been investigated by Waghmare\textit{ et al.},~\cite{Waghmare} who
reported the Born effective charge, LO-TO splitting, and the optical
dielectric constants. However, the dielectric properties of GeTe in the
stable ferroelectric phase have not been examined. Ciucivara\textit{ et al.}~\cite{Ciucivara} 
used \textit{ab initio} calculations reporting a value
of 10.11$e$ for the Born effective charge in the ferroelectric phase. Whereas such a
value is very close to the value previously reported for the rocksalt
structure,\cite{Zein} it is in clear contrast with the behaviour for perovskites where
large Born effective charge modifications due to the ferroelectric
transition have been reported.

The aim of our work was to perform a comprehensive study of the electronic,
dynamical, dielectric, elastic and piezoelectric properties of GeTe in its
low-temperature rhombohedral phase. Doing so, we uncover some problems with
previous calculations. The correct understanding of the bulk ferroelectric
phase of GeTe is a prerequisite for future investigations of the
ferroelectric properties of the more complex GeTe nanostructures.~\cite%
{Sun,Lee1}

This paper is organized as follows: In Sec. II, we present the
details of the methods used in the present study. In Sec. III, we
discuss the ground-state structural and electronic properties of
GeTe. In Sec. IV, we present the calculated Born effective charge
and optical dielectric tensors. In Sec. V and VI, the phonon band
structure and IR reflectivity are presented. In Sec.VII and VIII,
we present the elastic and piezoelectric properties.

\section{technical informations}

All the calculations have been performed using plane waves and
norm-conserving pseudopotentials, as implemented in the ABINIT code.~\cite%
{ABINIT, abinit05} The dielectric, dynamical, elastic and
piezoelectric properties have been
evaluated within the density-functional perturbation theory.~\cite%
{Baroni01,XGonze10337,XGonze10355,Xifan} We employed Hartwigsen-Goedecker-Hutter pseudopotentials,~\cite{HGH} 
generated including spin-orbit coupling, within the
local-density approximation adopting the Teter Pade parameterization.\cite{XC}%
Although this approximation is of frequent use, and gives
correct trends, for the study of dielectric materials, one should
be aware of some inherent limitations, due to the neglect of the
polarisation dependence, present in the ``exact''
functional.~\cite{GGG95,GGG97} The inclusion of spin-orbit coupling
does not lead to noticeable changes for most of the results
presented here, except for the electronic band structure, which
was already known.~\cite{Tung} Single-particle wave functions were
expanded using a plane wave basis up to a kinetic energy cutoff
equal to 15 Ha. The Brillouin zone integration
was performed using special $k$-points sampled within the Monkhorst-Pack scheme%
.~\cite{Monkhorst-Pack} Even though the electronic and structural properties
were found to be well converged using $4\times4\times4$ $k$ point mesh, a
denser mesh of $16\times16\times16$ $k$-points was required to describe well the
vibrational properties. For reasons of consistency, the latter grid was used
throughout this work for all calculations.

\section{structural and electronic properties}

The ground state rhombohedral structure of GeTe (space group
\textit{R3m}) has been represented in our calculations taking the
$z$ axis along the [111] of the conventional distorted rocksalt
structure, with primitive translation vectors $(\frac{a}{4}, -\frac{a}{4%
\sqrt{3}}, \frac{c}{3})$, $(0,\frac{a}{2\sqrt{3}},\frac{c}{3})$ and $(-\frac{%
a}{4},-\frac{a}{4\sqrt{3}}, \frac{c}{3})$, where the present $a$
and $c$ are
related to $a_0$ and $\alpha$ of Table \ref{structural} via the relations $%
a=2a{_0}\sin(\frac{\alpha}{2})$ and
$c=a_0\sqrt{3(1+2\cos(\alpha))}$. In Table~\ref{structural}, we
report the calculated structural parameters. The lattice
parameters $a_{0}$, the angle $\alpha $, and the shift of the Ge
and Te sublattices are reported using the distorted rocksalt
setting : atomic positions are Ge (0 0 0) and (0.5-$\tau $
0.5-$\tau $ 0.5-$\tau $) for Ge and Te respectively. Our results
are globally in good
agreement with previously reported \textit{ab initio} results.~\cite%
{Ciucivara}. Moreover the deviation between the calculated structural
parameters($a_{0}$,$\alpha $) and experiment is less than 2\%.

\begin{figure}[t]
\begin{center}
\epsfig{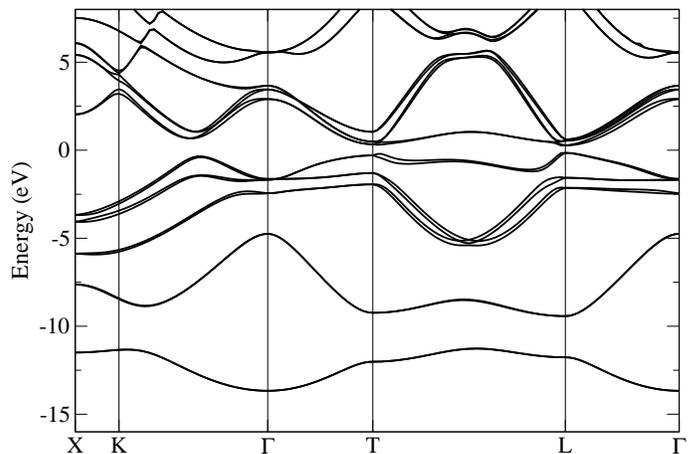}
\end{center}
\caption{Calculated band structure of GeTe}
\label{bandstructure}
\end{figure}

In Figure \ref{bandstructure} we show the energy band structure calculated
using the theoretical structural parameters from Table \ref{bandstructure}.
The general features of the band structure agree with previously reported
\textit{ab initio} calculations.~\cite{Rabe2,Ciucivara} The calculated
direct energy gap at L is 0.48 eV, this value is slightly larger 
than the previously reported \textit{ab initio} results of 0.40~\cite{Rabe2}
and 0.369 eV.~\cite{Ciucivara} The indirect gap exists at
(L,(0.35,0.2,0.2) ) is 0.28 eV slightly bigger than the tunneling
spectroscopy results 0.2 eV.~\cite{Chang} Since the electronic gap is quite sensitive 
to the distortions, it was suggested by Rabe\textit{ et al.} that such overestimation of the gap can be 
due to the difference between the theoretical structural parameters used in the 
calculations and those of the thin films used in the tunneling spectroscopy measurements.\cite{Rabe2}

\begin{table}[b]
\caption{{Calculated structural parameters of GeTe. The lattice parameter $%
a_0$ (in $\mathring{A}$), the angle $\protect\alpha$ (in deg), the deviation
from of the Ge sublattice from the 0.5 sublattice position $\protect\tau$,
and the volume $\Omega=(a^{3}_{0}/4)\sin^{2}\protect\alpha$ (in $\mathring{A}%
^3$). The calculated lattice constants $a$ and $c$ (in $\mathring{A}$) of the
equivalent hexagonal representation of the unit cell are also presented.} }
\label{structural}
\begin{center}
\begin{ruledtabular}
\begin{tabular}{c c c c c c cc    }
&  $a_0$   & $\alpha$& $\tau$ & $\Omega$    & $a$   & $c$  \\
        \hline\hline&&&\\
Present & 5.893 & 88.96 & 0.0236 & 51.15 & 8.258 &  10.391   \\
Theory~\cite{Ciucivara} & 5.886 &89.24  & 0.0217& 50.96      &   &   \\
Exp ~\cite{Onodera} &5.98 & 88.35  &  0.0248 &53.31  &    & \\
Exp ~\cite{Goldak} &5.996 & 88.18  &  0.026 &  &    & \\

\end{tabular}
\end{ruledtabular}
\end{center}
\end{table}

\section{dielectric properties}

We have first calculated the Born effective charge tensor of atom
$\kappa$, defined as the induced polarization of the solid along
the direction $i$ by a unit displacement in the direction $j$ of
the sublattice of atom $\kappa$ at vanishing electric field $E$  

\begin{equation}  \label{eqnzz}
Z^{*}_{\kappa,ij}=\Omega\frac{\partial P_i}{\partial a_{\kappa
j}}\biggr|_{E=0}
\end{equation}

Using Eq.~(\ref{eqnzz}), the Born effective charge tensor can be calculated
using DFPT or within the finite-difference method (FDM). In the latter case the
polarization is calculated using the Berry phase technique.~\cite{Vanderbilt}
~We have employed both methods in this study and the results are
shown in Table~\ref{tab:bo}. Due to the symmetry properties of
\textit{R3m} structure, the Born effective charge tensors of
Ge and Te are diagonal, with two independent components, along ($%
Z_{\parallel }^{\ast }$) and perpendicular ($Z_{\perp }^{\ast }$) to the
trigonal axis.

\begin{table}[h]
\caption{{Components of the Born effective charge tensor as calculated by
DFPT and FDM, and the calculated optical dielectric tensor
in units of vacuum permittivity $\protect\varepsilon_0$. } }
\label{tab:bo}
\begin{center}
\begin{ruledtabular}
\begin{tabular}{ccccccccccccccc}
                    & & \multicolumn{2}{c}{DFPT}
                    & & \multicolumn{2}{c}{FDM}
                    & & \\
\cline{3-4} \cline{6-7}
                   & & $Z^{*}_{\perp}$ & $Z^{*}_{\parallel}$ & & $Z^{*}_{\perp}$  & $Z^{*}_{\parallel}$ &
$\varepsilon^{\infty}_\perp$ & $\varepsilon^{\infty}_\parallel$ \\
\hline\hline&&&&&&&&&\\
                   & & 6.89  &4.69 && 6.99 &  4.72 & 59.43 & 49.23
\end{tabular}
\end{ruledtabular}
\end{center}
\end{table}

The Born effective charge tensor is strongly anisotropic with a difference of 2.2
between $Z^{*}_\parallel$ and $Z^{*}_\perp$ components. Both $Z^{*}_{\perp}$
and $Z^{*}_{\parallel}$ are significantly larger than the nominal ionic
value of $+2$ for Ge and $-2$ of Te.~\cite{Note} This is a consequence of
partial hybridization of the $p$ orbitals of both Ge and Te which results in
a mixed ionic-covalent bond. 

Our results for the Born effective charge of the ferroelectric
phase differs widely from the previous results obtained using 
the Berry phase and FDM.~\cite{Ciucivara} A value of $Z^{*}_\parallel=10.11$
was reported which is much larger than our calculated value 
of 4.72. However, the agreement between $Z^{\ast }$ 
from DFPT and FDM as clearly
seen in Table~\ref{tab:bo}, questions the accuracy of the results
obtained in Ref.~\onlinecite{Ciucivara}. 

The values of both $Z_{\parallel }^{\ast }$ and $Z_{\perp }^{\ast }$ are
also smaller than those theoretically obtained for rocksalt high temperature phase.~\cite{Zein,Waghmare}
It has been noted previously that, for perovskite systems, $Z^{\ast }$ is strongly dependent
on the geometry.~\cite{Ghosez95} To get more insight about the evolution of $%
Z^{\ast }$ due to the paraelectric-ferroelectric phase transition, we have
calculated $Z^{\ast }$ in the rocksalt structure, then manually displaced
atoms toward the ferroelectric positions $(\lambda )$ with small steps $%
(\delta \tau )$ calculating $Z^{\ast }$ at each step, keeping the unit cell
structure fixed at the high temperature \textit{F3m} lattice vectors.

\begin{figure}[tbp]
\begin{center}
\epsfig{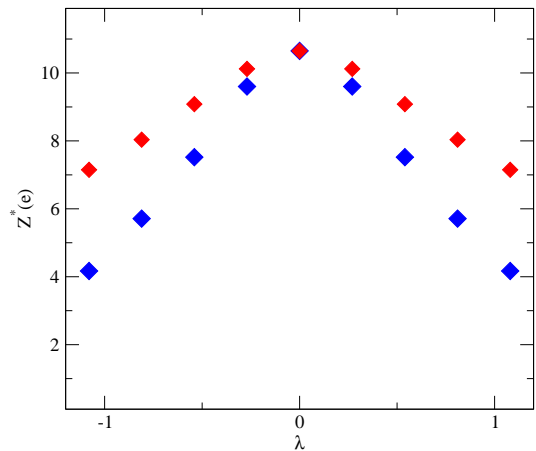}
\end{center}
\caption{(color online) Evolution of longitudinal Born effective charge $%
Z^*_\parallel$ (blue) and transverse Born effective charge $Z^*_\perp$ (red)
as a function of atomic displacement from the rocksalt ideal positions.}
\label{zz}
\end{figure}

As seen in Figure \ref{zz} , the Born effective charge in the rocksalt phase
is 10.68, in good agreement with the theortical results of Waghmare~\textit{ et al.},~\cite{Waghmare}
who reported a value of 10.8.  It has been shown previously that such high value of Born effective 
charge can be due to extremely narrow electronic gap possessed by 
the high temperature rocksalt phase.~\cite{Zein} We have
noticed that both $Z^*_\perp$ and $Z^*_\parallel$ drop as the atoms approach
the ferroelectric sites. Such modification of charge with atomic position is
considered a feature similar to what was obtained previously in case of ABO$%
_3$ perovskites.~\cite{Ghosez}

Similar to the effective charge tensor, the calculated dielectric
tensors are diagonal consisting of two independent components,
parallel and perpendicular to the trigonal axis. Our value is
larger than the previously reported experimental value (35-37.5).~\cite{Tsu} 
Generally speaking the high value of
($\varepsilon ^{\infty }$) comes as a consequence of the low value
of the electronic gap.

\section{Dynamical properties}

\begin{figure}[b]
\epsfig{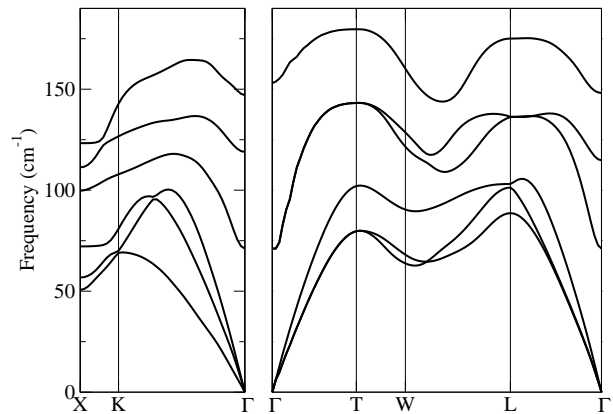} 
\caption{{Calculated phonon dispersion curves of ferroelectric GeTe at the
theoretical lattice parameters.} }
\label{phonon}
\end{figure}

Since there are two atoms per primitive unit cell, there will be
six phonon branches. The phonon branches are divided into three
acoustic and three optical phonon modes. Along the $\Gamma-T$
direction (trigonal axis), the phonon dispersion curves can be
classified as $E$ or $A_1$, according to whether the atomic
displacements are perpendicular or parallel to the trigonal axis,
respectively. Along the other directions, the branches cannot be
classified as pure $E$ or $A_1$ modes.

In Figure~\ref{phonon}, we show the calculated phonon band
structure obtained by DFPT. The general feature of the band
structure are close to those of Bi which has a similar structural
unit cell.~\cite{Sanchez} Because of the nonvanishing components of
the Born effective charge tensors, the dipole-dipole interaction
had to be properly included in the calculation of the interatomic
force constants.~\cite{XGonze10355} Such inclusion of the
dipole-dipole interaction in the interatomic force constants
results in splitting of longitudinal and transverse optic modes
(LO-TO splitting). Within this treatment, the $A_1$ and $E$ modes
(for wavevector aligned with the trigonal direction) are
characterized as LO and TO modes respectively. The crystal being
uniaxial, the LO-TO splitting at $\Gamma$ might vary with the
limiting direction. We actually notice a very strong dependence of
the LO-TO splitting on the wavevector.

It is difficult to grow large high quality GeTe crystals, so,
little experimental information on its vibrational modes has been collected.
The comparison between the calculated mode frequencies at $\Gamma $ with
those measured by Raman scattering experiments~is also difficult.~\cite{Andrikopoulos,Steigmeier} 
It is quite known that GeTe samples are
non-stoechiometric and contain a high concentration of free holes.~\cite{Bahl}%
 This results in complete screening of the dipole-dipole
interaction by the conduction electrons at zone
center.~\cite{Steigmeier} However, the comparison can be possible
if we artificially enforce such a complete screening of the
dipole-dipole interaction at the zone center. The results are
shown in Table~\ref{phonong}, the first row contain the values of
frequencies as calculated by including the dipole-dipole
interaction using a semiconducting screening, the second row
contained the values of frequency as calculated by complete
screening of the dipole-dipole interaction.~\cite{footnote} We note
that the values calculated by complete screening of the long range
force are in more agreement with the existing Raman inelastic scattering results.~\cite{Andrikopoulos,Steigmeier} 
Note that in absence of the LO-TO splitting as a result of complete screening of
dipole-dipole interaction, the $A_{1}$ mode has been characterized
experimentally as a TO mode. In fact, it was suggested that the
softening of this mode at the critical
temperature is responsible for the ferroelectric-paraelectric transition%
.~\cite{Steigmeier,footnote}

\begin{table}[tbp]
\caption{{Phonon frequencies at the zone center (in cm$^{-1}$) calculated for a
$\mathbf{q}$ vector $\parallel$ to trigonal axis.} }
\label{phonong}
\begin{center}
\begin{ruledtabular}
\begin{tabular}{ c c c }
             &      $E(TO)$  & $A_1(LO)$       \\
\hline\hline\\
Semiconducting    &      71  &  153 \\
Complete screening   &     100  &  134 \\
Exp\cite{Steigmeier}  &      98    &  140\\
Exp\cite{Andrikopoulos}&    80  &  122       \\

\end{tabular}
\end{ruledtabular}
\end{center}
\end{table}

\section{IR spectroscopy}

\begin{table}[b]
\caption{{Calculated values of IR oscillator strength tensor $%
S~(\times10^{-5})$ (in a.u.) of optical modes and components of static
permittivity tensor in units of $\protect\varepsilon_0$. } }
\label{IR}
\begin{center}
\begin{ruledtabular}
\begin{tabular}{ c c c c c c }
             &      $\perp$  &      $_{\parallel}$ \\
\hline\hline\\
 $S(E)$               &      ~5.74,~50.45 & 0.00\\
 $S(A_1)$             &        0.00 & 26.04\\
  \hline
 $\varepsilon(0)$   &       254.89     & 81.58 \\

\end{tabular}
\end{ruledtabular}
\end{center}
\end{table}

The dielectric tensor $\varepsilon (\omega )$ in the lowest frequency range
can be related experimentally to the IR spectra. It can be calculated
theoretically by accounting for ionic relaxations in the calculations of the
permittivity tensor. The ionic contribution to $\varepsilon
(\omega )$ comes mainly from optical phonon contributions (without damping)
to the IR oscillator strength $S_{m,ij}$

\begin{equation}
\varepsilon_{ij}(\omega)=\varepsilon^{\infty}_{ij}+\frac{4\pi}{\Omega}%
\sum_{m}^{TO}\frac{S_{m,ij}}{\omega^{2}-\omega^2}
\end{equation}

where $\Omega$ is the volume of unit cell and $m$ is the phonon mode rank.

In Table \ref{IR}, we present the calculated IR oscillator
strength and the components of $\varepsilon (0)$ for a perfect
semiconductor GeTe crystal.
From Table \ref{IR}, it is obvious that the ionic contribution to $%
\varepsilon (0)_{\perp }$ comes purely from $E$ modes and to $\varepsilon
(0)_{\parallel }$ from $A_{1}$ mode. The inclusion of the ionic contribution
results in a strongly anisotropic $\varepsilon (0)$ with $\varepsilon
(0)_{\perp }$ almost three times larger than $\varepsilon (0)_{\parallel }$.
In fact, the strong anisotropy of $\varepsilon (0)$ has been
expected in regard of the  frequency difference between $E$(71 cm$^{-1}$) and $%
A_{1}$(119 cm$^{-1}$) modes. Another reason for the large
anisotropy comes from the fact that one of $S_{\perp }$ of $E$
modes is almost twice $S_{\parallel }$ of $A_{1}$ mode.

\begin{figure}[tbp]
{b}
\par
\begin{center}
\epsfig{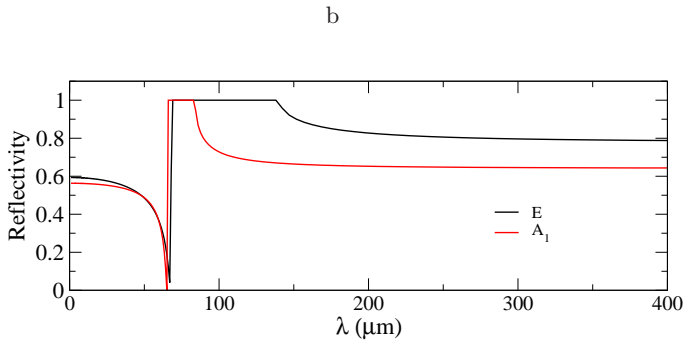}
\end{center}
\caption{(Color online) Calculated IR reflectivity spectra of GeTe (without
anharmonic damping). }
\label{refl}
\end{figure}

In Fig. \ref{refl} we present the calculated IR reflectivity associated with
$E$ and $A_{1}$ modes. The reflectivity related to $A_{1}$ mode can be
associated with light incident parallel to the trigonal axis of
the crystal i.e., perpendicular to (001) surface. Similarly, the $E$ modes
reflectivity shall be associated with light incident perpendicular to 
the trigonal axis, i.e., parallel to $a$ ($b$) directions which are
perpendicular to 100 (010) surfaces.

The range of light wavelength at which the maximum reflectivity occurs
differs depending on the surface. For example, whereas the maximum
reflectivity for light incident on (001) surface occurs for
wavelengths in the range 65-83 $\mu $m, it is in the range of 68-137 $%
\mu $m in case of 100 or 010 surfaces.

\section{Elastic properties}

In this section we present our results for the elastic and compliance
tensors. The elastic tensor is defined as the change in the stress of the
solid in the direction $\alpha$ as the strain is changing in the direction $%
\beta$, where $\alpha,\beta=1...6$ in Voigt notation

\begin{equation}
c_{\alpha\beta}=\frac{\partial \sigma_{\alpha}}{\partial \eta_{\beta}}
\end{equation}

The above equation split into two main contributions

\begin{equation}
c_{\alpha\beta}=\frac{\partial \sigma_{\alpha}}{\partial \eta_{\beta}}\biggr|%
_{u}+ \sum_k\frac{\partial \sigma_{\sigma}}{\partial u_{k,i}}. \frac{%
\partial u_{k,i}}{\partial \eta_{\beta}}
\end{equation}

The first term is the frozen (clamped) ion elastic tensor ($c^{0}$), the
second term includes contributions from force-response internal stress and
displacement-response internal strain tensors. The second term accounts for
the ionic relaxations in response to strain perturbations. The addition of
the two contributions is the relaxed ion elastic tensor $c$. The compliance
tensors is simply defined as the inverse of the elastic tensor.

In Table~\ref{elastic}, we show our results for the elastic and
compliance tensors. Due to the low symmetry of \textit{R3m} of
rhombohedral phase of GeTe, there are six independent elastic
constants. The obtained values for the elastic tensor constants
satisfy the mechanical stability restrictions
for trigonal type unit cells $c_{11}-|c_{12}|>0$, $%
(c_{11}+c_{12})c_{13}-2c_{13}^{2}>0$, $(c_{11}-c_{12})c_{44}-2c_{14}^{2}>0$.~\cite{Born} 
Usually, the inclusion of internal relaxations
reduces the values of elastic tensor components due to the relief of stress
tensor components. The same behavior is also reflected in the increase of
the compliance tensor components. The effect of internal relaxation is more
pronounced in the case of $c_{33}$ which represents the axial shear along
the surface perpendicular to the three-fold rotation axis. However, one
should emphasize here that the effect of internal atomic
relaxation is not as strong as for other materials such as ZnO and BaTiO$_{3}$%
.~\cite{Xifan} The small differences between the clamped and relaxed
compliance tensor might explain the similarity of the bulk modulus of the
ferroelectric phase and the rocksalt phase as has been suggested by the
experimental findings.~\cite{Onodera}

\begin{table}[htp]
\caption{{Clamped ion ${c}^0$ and relaxed ion $c$ elastic tensors (GPa),
clamped ion $s^0$ and relaxed ion $s$ compliance tensors (in TPa$^{-1}$)}. }
\label{elastic}
\begin{center}
\begin{ruledtabular}
\begin{tabular}{c c c c c }
index & ${c}^0$ &  ${c}$  & $s^0$ & $s$ \\
 \hline\hline
 11           &      116.16 &  112.74 &  11.66  &  12.28 \\
 12           &       22.34 &   19.84 &  -2.35  & -2.92 \\
 13           &       36.78 &   27.56 &  -3.94  & -4.31 \\
 14           &       27.08 &   24.68 &  -5.82  & -8.45 \\
 33           &       86.98 &   59.88 &  14.83  & 20.66 \\
 44           &       65.29 &   44.37 &  20.14  & 31.93 \\
\end{tabular}
\end{ruledtabular}
\end{center}
\end{table}

The bulk modulus can be readily calculated, using the above results, from
the compliance tensor,~\cite{Nye}

\begin{equation}
B=\left[\sum_{\alpha\beta}^3 s_{\alpha\beta}\right]^{-1}
\end{equation}

Using the above equation, we get a value of $B=45.17$ GPa which is in quite
good agreement with the experimental value of 49.96$\pm$3.2 GPa.~\cite{Onodera}

We have also calculated the bulk modulus by fitting the total energy as a
function of volume curve using the Murnaghan equation of state.
The obtained value of $B=44.3$ GPa is in much better agreement with
experimental findings than the theoretical value 
reported in Ref.~\onlinecite{Ciucivara}. 

Unfortunately we are not aware of any experimental result for the
elastic constants of GeTe. The agreement between the value of the
bulk modulus value extracted from the
elastic tensor and those calculated by fitting to the
equation of state or measured in experiment nevertheless insures
the overall reliability of the calculated values of the elastic tensor.

\section{Piezoelectric properties}

The proper pizeoelectric tensor $e$ is defined as the induced polarization
in $i$ direction due to a strain change for index $\alpha$

\begin{eqnarray}
e_{i\alpha}&=&\frac{\partial {P_{i}^\prime}}{\partial {\eta_{\alpha}}} \\
&=&\frac{\partial {P_{i}}^\prime}{\partial {\eta_{\alpha}}}\biggr|_{u}
+\sum_{k}\frac{\partial {P_{i}}^\prime}{\partial {u_{ik}}} .\frac{\partial {%
u_{ik}}}{\partial {\eta_{\alpha}}}
\end{eqnarray}

Where $P{^{\prime }}_{i}$ is the reduced (rescaled) polarization as defined
in the appendix of Ref.~\onlinecite{Xifan}. The first term is the proper
homogeneous strain contribution to the piezoelectric tensor $%
e^{0}$ which arises mainly from the sole  electronic
contribution. The second term, often called internal strain piezoelectric
tensor, includes contributions from the Born effective charge tensor
diagonal and internal relaxation. The second term represents the ionic
contribution to piezoelectric tensor.

\begin{table}[htb]
\caption{{Independent components of proper homogeneous piezoelectric tensor $%
e^0$, internal strain piezoelectric tensor, and the proper total
piezoelectric tensor (in C/m$^2$). The piezoelectric constant tensor $d$
(in pC/N) and the electromechanical coupling constants $k$ are also given}. }
\label{piezo}
\begin{center}
\begin{ruledtabular}
\begin{tabular}{c c c c c c }
                       & 15       & 21    & 31        & 33\\
\hline
 $e^0$                      &   ~0.88  & 1.24    & 0.89      & --0.31\\
 internal strain       &  --6.01  &--0.69  & --0.95      &  --2.78\\
\hline
 Total                 &  --5.13  &0.55  & --0.06       &  --3.09\\
\hline
$d$                     & --173.00  &~51.68  & ~12.78  &  --63.44\\
$k$                     & 0.54  & 0.26 &  0.12  &  0.49\\
\end{tabular}
\end{ruledtabular}
\end{center}
\end{table}

In Table~\ref{piezo}, we present the results for the piezoelectric
tensor. There are 4 independent components. The components
$e_{31},e_{33}$ represent the induced polarization along the
trigonal axis created in response to shear strain in the
$ab$-plane and along the trigonal axis respectively. The
other components describe the induced polarization along the primitive axes (%
$a$,$b$) by shear strain. The calculated value of $e_{33}$ is --3.09 (C/m$%
^{2}$). The proper homogeneous strain contribution has been found to be
--0.31(C/m$^{2}$), however the strain contribution is much larger, adding
--2.78 (C/m$^{2}$). The strain contribution was relatively large in the case
of $e_{15}$ with a value of --6.01 (C/m$^{2}$). The strain contribution of
the last case being twice the value of $e_{33}$ can be explained by the
noticeable anisotropy in the Born effective charge, and by the large
strained-induced ionic motion in the lateral direction in response of a
strain applied along the trigonal axis. For the other elements, the strain
contribution reduces the polarization, which is almost cancelled, as in the case of $e_{31}$.

The efficiency of the produced electric energy vs the spent mechanical
energy can be estimated by calculating the electromechanical coupling
constant, defined as
\begin{equation}
k_{i\alpha}=\frac{|d_{i\alpha}|}{\sqrt{\varepsilon^{\sigma}_{ii}S_{\alpha%
\alpha}}},
\end{equation}
where $d_{i\alpha}$ is the piezoelectric constant $d_{i\alpha}=S_{\alpha%
\beta}e_{i\beta}$, and $\varepsilon^{\sigma}$ is the free stress dielectric
tensor, related to the vanishing strain dielectric tensor shown in Table ~%
\ref{piezo} via Eq.~20 of Ref.~\onlinecite{Xifan} .

Similar to the zero strain dielectric tensor $\varepsilon$, the stress free dielectric 
tensor $\varepsilon^{\sigma}$ is
diagonal, with two independent values $\varepsilon^{\sigma}_\parallel$ =
90.70$\varepsilon^0$ and $\varepsilon^{\sigma}_\perp$=360.09$\varepsilon^0$.
Note $\varepsilon^{\sigma}$ have components larger than $\varepsilon$ as
expected for any piezoelectric material.

In the last two rows of Table~\ref{piezo}, we show our calculated
piezoelectric constants $d$ and electromechanical coupling constants. Even
though the various components of $d$ are quite smaller than those reported
for PMNT system,~\cite{Wang} it is interesting that our calculated value of $%
d_{15}$ is comparable with that of the giant piezoelectric materials PMN-PT
and PZN-PT~\cite{Hu} where $d_{15}$ extends between 131-190 pC/N. On the
other hand, the calculated electromechanical coupling factors in general are
far less compared with that of PMN-PT and PZN-PT~\cite{Hu} or PMNT\cite{Wang}%
, however they are slightly better than those of ZnO.~\cite{Xifan}

\section{Conclusion}

We have investigated the dielectric, dynamical and mechanical properties of
the ferroelectric phase of GeTe within density-functional perturbation
theory. Our study covers all the linear couplings between applied static
homogeneous electric field, strain, and periodic atomic displacements : Born
effective charge, dynamical matrix at the zone-center, clamped and dressed
elastic constant, optical dielectric tensor, adiabatic dielectric tensor,
free stress dielectric tensor, piezoelectric coefficients, elastic and
compliance tensor. We also examined the phonon band structure, as well as
the change of the Born effective charge tensor with the atomic positions.
Concerning the latter, our results disagree with those of Ciucivara\textit{ et al.}~\cite%
{Ciucivara} by a factor of two. Since we observe a rapid variation
of the Born effective charge with the atomic positions, we
hypothetize that their result has not been obtained for the
ferroelectric phase, but for the symmetric paraelectric phase. Our
other results have also been discussed, and compared with the
available theoretical and experimental results.

\begin{acknowledgments}
We acknowledge financial support by the Interuniversity Attraction
Poles Program (P6/42) - Belgian State - Belgian Science Policy.
Two of the authors (R.S. and X.G.) acknowledge support from the
the Communaute Francaise de Belgique (Action de Recherches
Concertee 07/12-003) and the European Union (NMP4-CT-2004--500198,
``NANOQUANTA'' Network of Excellence ``Nanoscale Quantum
Simulations for Nanostructures and Advanced Materials'', and
``ETSF'' Integrated Infrastructure Initiative) and FAME-EMMI
Network of Excellence "Functionalized Advanced Materials
Engineering".
\end{acknowledgments}

\bibliographystyle{plain}
\bibliography{basename of .bib file}

\begin{thebibliography}{99}
\bibitem{Note} The nominal ionic value of Ge in tetrahedral covalent
crystals is +4. However in binary compounds as IV-VI, it is considered +2
which is the negative of the nominal ionic value of Te.
\bibitem{footnote} In practice, the screening of the depolarizing
field associated to the A1(LO) mode will reduce its frequency to
that of A1(TO) mode.
\end{thebibliography}

\end{document}